\documentclass[10pt, letter]{article}

\usepackage[dvips]{epsfig}

\begin{document}

\author{N. Scafetta$^{1}$, P. Grigolini$^{2,3}$, P. Hamilton$^{4}$  and B. J. West$^{1,5} $ \\
1) Physics Department, Duke University, Durham, NC\\
2) Physics Department, University of North Texas,  Denton, TX\\ 
3) Dipartimento di Fisica dell'Universit\`a di Pisa and   INFM, \\Pisa 56127, Italy\\
4) Center for Nonlinear Science, Texas Woman University, \\Denton, TX 76204 \\
5)  Mathematics Division, Army Research Office, \\Research Triangle Park, NC}

\title{Non-extensive diffusion entropy analysis:\\ non-stationarity in  teen birth phenomena}
\maketitle

   \begin{abstract}
A complex process  is often a balance between  non-stationary and  stationary components. We show how the non-extensive Tsallis q-entropy indicator may be interpreted as a measure of non-stationarity in time series. This is done by applying the non-extensive entropy formalism to the Diffusion Entropy Analysis (DEA). 
We apply the analysis to the 
study of the teen birth phenomenon. We find that the unmarried 
teen births are strongly influenced by social processes with memory. This memory is related to the strength of the non-stationary component of the signal and is 
more intense than that in the married teen time series. By using the wavelet multiresolution analysis we attempt to give a social interpretation of this effect.

 \end{abstract}

\section{Introduction}

One of the most exciting and rapidly developing areas of modern research is the quantitative study of ``complexity." Complexity has special interdisciplinary impact in the fields of physics, mathematics, information science, biology, sociology and medicine. No definition of complex system has been universally embraced, so here we adopt the working definition ``an arrangement of parts so intricate as to be hard to understand or deal with." Therefore, the main goal of the science of complexity is to develop mathematical methods able to discriminate among the fundamental microscopic and macroscopic constituents of a complex system and to describe their interrelations in a concise way.    
   
Experiments usually yield results in the form of time series for physical observables. Typically these time series contain both a slow regular variation, usually called ``signal," and a rapid erratic fluctuation, usually called "noise." Historically the techniques applied to  processing such time series have been based on equilibrium statistical mechanics and, therefore, they are not applicable to phenomena  far from equilibrium. 
Among the most egregious of the limitations of these popular methods are the assumptions of stationarity of the two time correlation function and Markovian memory. 

In this paper we show that the non-extensive Tsallis q-entropy indicator may be interpreted as a measure of the strength of the non-stationary component of a time series. This is done by applying the non-extensive entropy formalism to  Diffusion Entropy Analysis (DEA). DEA is a recent and very efficient method developed to detect the scaling of a stationary complex process; The scaling of the probability density function (pdf) of the diffusion process generated by time series imagined as a physical source of fluctuations, see Ref. \cite{scafetta,grigolini,dea3,scalingdetection,compressionalgorithm,nicola}. 
 
We apply the above analysis to the study of the teen birth phenomenon. The daily birth data cover the number of births to married and unmarried teens in Texas during the period  1994 to 1998. Time series analysis in the social sciences is traditionally done using
linear models, such as analysis of variance and linear regression.
Underlying these techniques is the assumption that the phenomena of
interest, such as adolescent sexuality, pregnancy and other developmental
processes, are stationary, \cite{rodgers}. However, this is not a comprehensive  approach  because the births by teenagers are characterized by a complicated annual cycle that is the  source of non-stationarity.  
We find that  unmarried teens seem to be more strongly influenced by social processes
  than are the married teens. Finally, by using the wavelet multiresolution analysis we attempt to give a social interpretation of this effect.

\section{Non-extensive diffusion entropy analysis and its non-stationary meaning}

DEA detects the scaling  of a stationary process through the study of the time evolution of the Shannon entropy of the pdf in terms of the diffusion process generated by the time series. 
A time series $\{\xi_i\}$ may be interpreted as diffusion fluctuations. As in a random walk, we define  diffusion trajectories  by
        \begin{equation}
    x^{(z)}(t) =  \sum_{i = 1}^{t} \xi_{i+z}~,
        \label{positions}
        \end{equation}
where $z=1,~2, ~3, ...~$.
These trajectories generate a diffusion-like process that is described by
a pdf $p(x,t)$, where $x$ denotes the  variable collecting the fluctuations and $t$ is the diffusion time. 
The pdf of a stationary diffusion process is  expected to have the fundamental scaling property
\begin{equation}\label{stationarycondition}
p(x,t) =  \frac{1}{t^{\delta}}~F\left( \frac{x}{t^{\delta}}\right)~,
\end{equation}
where the coefficient $\delta$  is the scaling exponent.

The Shannon entropy is defined by
\begin{equation}
S(t) = - \int_{-\infty}^{+\infty} dx ~p(x,t) \ln[p(x,t)]~.
\label{diffusionentropy}
\end{equation}
Using the stationary condition of Eq.(\ref{stationarycondition}) we
obtain
\begin{equation}
S(t) = A + \delta~ \ln(t),
\label{linearincrease}
\end{equation}
with
\begin{equation} \label{A}
A \equiv -\int_{-\infty}^{\infty} dy \, F(y) \, \ln [F(y)]~,
\end{equation}
where  $y = x/t^{\delta}$. 
Eq.  (\ref{linearincrease}) indicates that in the case of stationarity,
the entropy $S(t)$ increases linearly with $\ln(t)$. Numerically, the scaling exponent $\delta$ can be evaluated by using fitting curves with function of  the form $f_S(t)=K + \delta \ln (t)$ that, when graphed on linear-log graph paper yields  straight lines.

The breakdown of the stationary condition may be simulated by assuming that the scaling exponent $\delta$ of Eq. (\ref{stationarycondition}) changes with time. This can be implemented by assuming Eq. (\ref{stationarycondition}) has the non-stationary general form  
\begin{equation}
p(x,t) =  \frac{1}{t^{\delta(t)}}~F\left( \frac{x}{t^{\delta(t)}}\right)~.
\label{nonstationarycondition}
\end{equation}
If we assume that
\begin{equation}
\delta(t) = \delta_{0} + \eta \ln(t), 
\label{logarithmicchange}
\end{equation}
where $\delta_{0}$ and $\eta$ are two constants,  we notice that, in the new
non-stationary condition, the traditional entropy (\ref{diffusionentropy}) yields:
\begin{equation}
S(t) = A + \delta_{0} \ln(t) + \eta ~[\ln(t)]^{2}.
\label{quadratic}
\end{equation}
The quadratic form of Eq. (\ref{quadratic}) suggests  that the choice of $\delta(t)$ given by Eq. (\ref{logarithmicchange}) has the mathematical meaning of the quadratic term in the Taylor expansion of the diffusion entropy (\ref{diffusionentropy}). As a consequence, we should expect that, in general, $\delta(t)$ always assumes  the form of Eq. (\ref{logarithmicchange}), at least for small values of $\ln(t)$. 

Let us see how all this may be related to the non-extensive Tsallis q-indicator \cite{tsallis}. The Tsallis non-extensive entropy  reads
\begin{equation}
S_{q}(t) = \frac{1 - \int_{-\infty}^{+\infty}dx ~p(x,t)^{q}}{q-1}.
\label{nonextensiveentropy}
\end{equation}
It is straightforward to prove that this entropic indicator coincides
with that of Eq.(\ref{diffusionentropy}) in the limit where the entropic index $q \to 1$. Let us make the assumption that in 
the diffusion regime the
departure from this traditional value is weak and assume  $\epsilon \equiv q - 1\ll 1$.
This allows us to use the following approximate expression for the
non-extensive entropy
\begin{equation}
S_{q}(t)  =   - \int_{-\infty}^{+\infty} dx ~p(x,t) \ln[p(x,t)]    -  \frac{\epsilon}{2}  \int_{-\infty}^{+\infty} dx ~p(x,t) \ln^{2}[p(x,t)]. 
\label{firstorder}
\end{equation}
In the specific case where the non-stationary condition of
Eq.(\ref{nonstationarycondition}) applies, this entropy yields the form
\begin{equation}
S_{q}=A - \epsilon B +(1- \epsilon A) \delta(t) \ln(t)- \frac{\epsilon}{2} 
~\left[\delta(t) \ln(t)\right]^{2},
\label{explicit}
\end{equation}
where $A$ and $B$ are two constants related to $F(y)$ of Eq. (\ref{nonstationarycondition}).
The regime of linear increase in $\ln(t)$ is recovered when $\epsilon$ is assigned the value
\begin{equation}
\epsilon = q-1= \frac{\eta}{\delta_{0}^{2}/2  + \eta A}~.
\label{crucialvalue}
\end{equation}

These theoretical remarks demonstrate that the non-extensive approach to
the diffusion entropy makes it possible to detect the strength of the
deviation from the stationary condition. In fact,
Eq.(\ref{crucialvalue}) establishes that $\epsilon= q-1 = 0$ implies $\eta=0$ that, according to Eq. (\ref{logarithmicchange}), indicates the
stationary condition. 
The conclusion of this Section is that the breakdown of the stationary
property  of Eq.(\ref{stationarycondition}) can be revealed by the
DEA under the form of an entropic index $q$
 departing from the condition of ordinary statistical
mechanics, namely $q = 1$. However, the breakdown of the stationary
property in a diffusion process will not last for ever. The Central Limit Theorem \cite{gardiner} states that for $t \to \infty$ the diffusion pdf converges to a stationary Gaussian form. Therefore, we expect, for  large $t$, to recover the stationary condition that is manifest when $q=1$. 
      
Fig. 1 shows the effect of  the non-extensive Tsallis q-entropy indicator as a function of time $t$ applied to the binomial distribution generated by a simple random walk:
 $q=1$ (solid line), $q=1.2$ (dotted line) and $q=0.8$ (dashed line).     The figure shows clearly the relation between $q$ and the bending shape  of the entropic curve, typical of the non-stationary condition expressed by Eq. (\ref{quadratic}).  Of course, in the case of a random walk, the  linear increase in $\ln(t)$ of the entropy $S_q(t)$ is recovered when $q=1$ and the scaling exponent is  $\delta=0.5$. This is the value of the scaling exponent  of the Gaussian distribution to which the binomial distribution  converges after few diffusion steps. 

A curiosity: What  happens if we adopt the R{\'e}nyi entropy \cite{beck} instead of the Tsallis entropy? It is easy to prove that the R{\'e}nyi q-entropy indicator has a simple parallel shifting effect instead of a bending effect upon the diffusion entropy and, therefore, it is not useful for our goal.

\section{The teen birth phenomenon analysis}

Texas is second only to California in the number of births to teens in the United States. Rates of birth to teens of all ages and racial/ethnic groups have been dropping in the United States since 1990, \cite{Ventura}. However, the size of the problem in Texas remains significant.
In 1996, in Texas there were 80,490 pregnancies and 52,273 births to girls 15-19 years old, \cite{national}. The U.S. rate of pregnancy among young women 15 to 19 years old was 97 per 1000 girls of that age, the rate in Texas was 113 per 1000. The mean age of teens giving birth was 17.62 years in Texas. Approximately 66\% of  births to teenagers in Texas were out of wedlock and 24\% of births to teens were to girls who had given birth at least once previously.
Data for the study reported here were abstracted from birth certificates obtained from the Texas Department of Health. The original time series was constructed from the daily count of births from January 1, 1994 through December 31, 1998. Every recorded birth to a woman under the age of 20 was included. Data on the marital status of the mother allowed us to analyze married and unmarried births separately. Reliable and valid birth certificate information regarding marital status did not become available in Texas until January 1, 1994. More information about the data may be found in Ref. \cite{scafetta}. 

The DEA of the data is preceded by a preliminary detrending to free the data from easily understandable linear and cyclical deterministic trends. In fact, the data show  a slight linear decrease (for married teen, Fig. 2a) and increase (for unmarried teen, Fig. 2b) trends and two strong periodicities: the annual trend due to the seasonal cycle and the weekly cycle due to social organization of the week into workdays and weekends. The weekends consistenly record a lower rate of births.  We eliminate the data associated with weekends and holidays, and detrend the linear ramp and annual frequency through the fitting curve:
\begin{equation}
\Xi(t) = A + B t + C \cos (\omega t ) + D \sin(\omega t).
\label{fittingcurve}
\end{equation}
In the case of the unmarried teens the fit gives  A = 97.5, B = 0.00893, C = 1.29, D = -6.30 and $\omega = 2 \pi/365.25.$  In the case of the married  we set A = 57.8, B 
= -0.00353, C = - 0.277, D = -4.14 and $\omega = 2 \pi/365.25.$   Figs. 2a and 2b show the original data as well as the detrended ones for the two groups.

Before applying the diffusion entropy algorithm to the two detrended datasets, we dichotomize the two signals, and associate the positive values to +1 and the negative values to -1. In this way, an easy confrontation with the random walk theory is possible. In fact,  if the new dichotomous series of +1 and -1 is random, the diffusion produced by its walks  gives the binomial distribution of the random walk that corresponds to the stationary condition with $q=1$ and $\delta=0.5$.  By the other side, if the new dichotomous series is not completely random but modulated by some type of memory, the correspondent diffusion process  shows some type of non-stationary behavior and we expect $q\neq 1$.   

Let us apply the non-extensive DEA to the two dichotomous detrended datasets. First, we build the diffusion trajectories according to the prescription of  Eq. (\ref{positions}). Second, as done in the random walk model, we calculate the probabilities/frequencies $p_i(t)$ that a trajectory occupies the $i^{th}$ position  at the diffusion time $t$. Finally,  
we evaluate the discrete non-extensive Tsallis entropy 
\begin{equation}
S_{q}(t) = \frac{1 - \sum_{i}p_i(t)^{q}}{q-1}
\label{nonexten2}
\end{equation}
and we look for the {\it magic} $q=Q$ that makes $S_{q}(t)$ increase linearly in $\ln(t)$,
at least in the first decades of the diffusion steps.  Figs. 3a and 3b show the diffusion entropy curves for $q=1$ and for $q=Q$. For married teens we get a magic Q close to 1; this means that the correspondent dichotomous series is random. Instead, for unmarried teens  we get $Q=1.257$ that reveals a non-stationary diffusion process and, therefore, a memory component in the signal. 

To investigate the nature of the memory  in the unmarried teen data, left after the detrending of the linear and  cyclical annual trends, we use the wavelet analysis \cite{percival}. The wavelets are a powerful method of analysis that localizes a signal simultaneously in time and frequency.  With a judicious use of the multiresolution wavelet transform, as explained in detail in Ref. \cite{nicola}, we can obtain an approximated distribution of  conceptions which result in birth during the year for both married and unmarried teenagers.  The estimated errors are $\pm2$ births against $\pm 2$ weeks. Moreover, we point out that identifying conception distribution from delivery dates among teens may be imprecise because of the high number  of miscarriages or abortions, almost $\% 49$ of  conceptions, and a sensitive seasonal dependency pre-term delivery in teens.  
Fig. 4 shows our  estimation of the daily number of conceptions, relative to the annual mean  value, for married (``+") and unmarried (``x") teenagers. The standard error is  2 births.   The conception rate in married teens changes regularly following the annual seasonal temperature cycle with a higher rate during the cold months and a lower rate during the hot months. Instead, the conception rate in unmarried teens seems more strongly influenced by 
the school-holiday yearly calendar. For example, there is a sharp drop of conceptions during the summer due probably to the fact that the schools are closed and the interactions with other teenagers is greatly reduced. It is the complex social component of the births to unmarried teens that is detected as non-stationarity by the DEA.

\section{Conclusion}

The new technique of analysis, based on the entropy of diffusion 
process, and consequently called DEA, has been proved by Refs. \cite{scafetta,grigolini,dea3,scalingdetection,compressionalgorithm,nicola} 
to be a very efficient method of scaling detection, which ensures the 
possibility of measuring the correct scaling coefficient $\delta$ when 
the property of Eq. (\ref{stationarycondition}) applies. This paper shows that the adoption 
of the Tsallis entropy rather than the Shannon entropy, as an entropy 
measure of the diffusion process, allows us to interpret the 
deviation of the Tsallis index q from the ordinary value $q = 1$, as an 
indicator of memory strength.

\newpage

\begin{figure}[h]
\epsfig{file=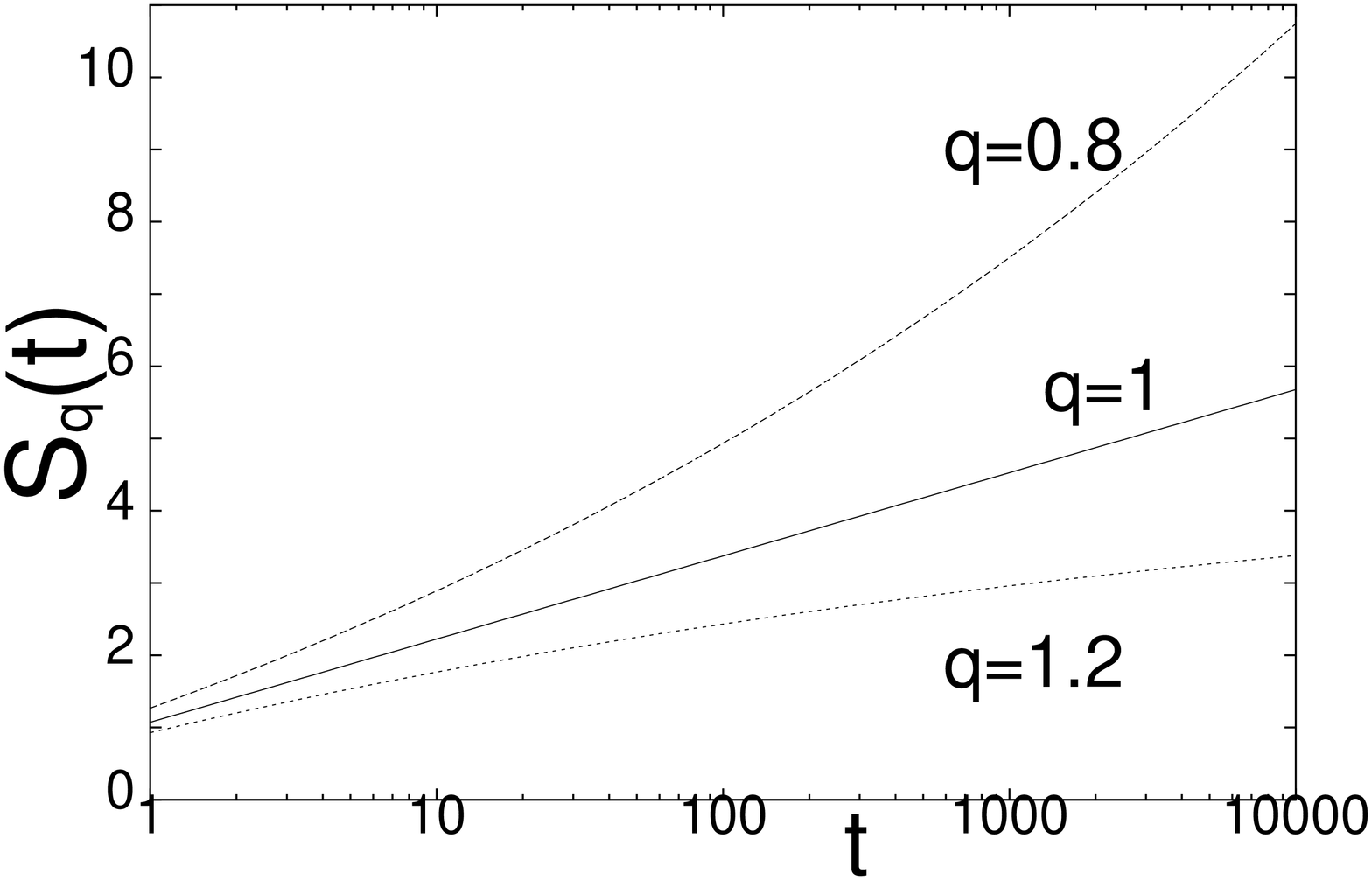, height=14cm,width=10cm,angle=-90}
\caption{The non-extensive Tsallis entropy as a function of time $t$ applied to the binomial distribution generated by the random walk:
 $q=1$ (solid line), $q=1.2$ (dotted line) and $q=0.8$ (dashed line).  }
\end{figure}

\newpage

\begin{figure}[h]
\epsfig{file=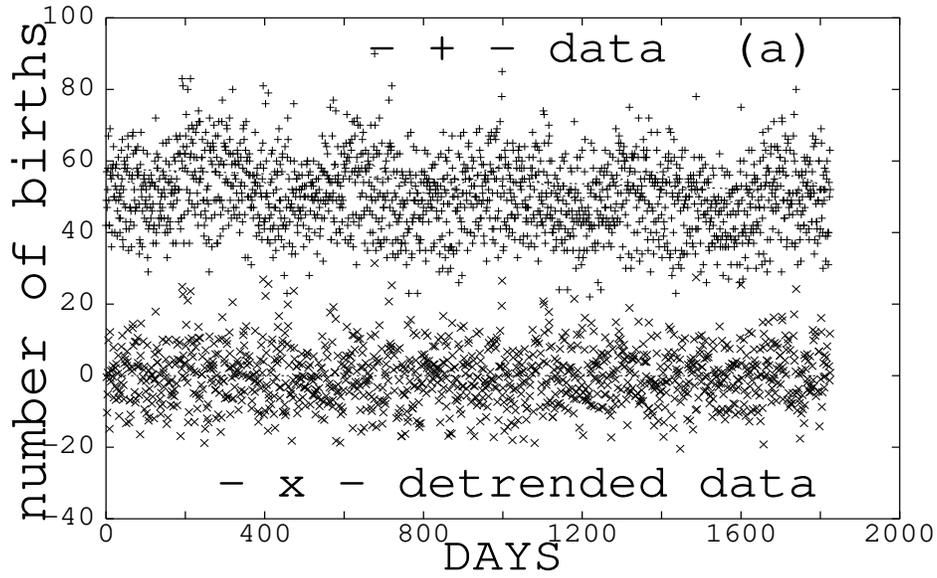, height=12cm,width=8cm,angle=-90}\\
\epsfig{file=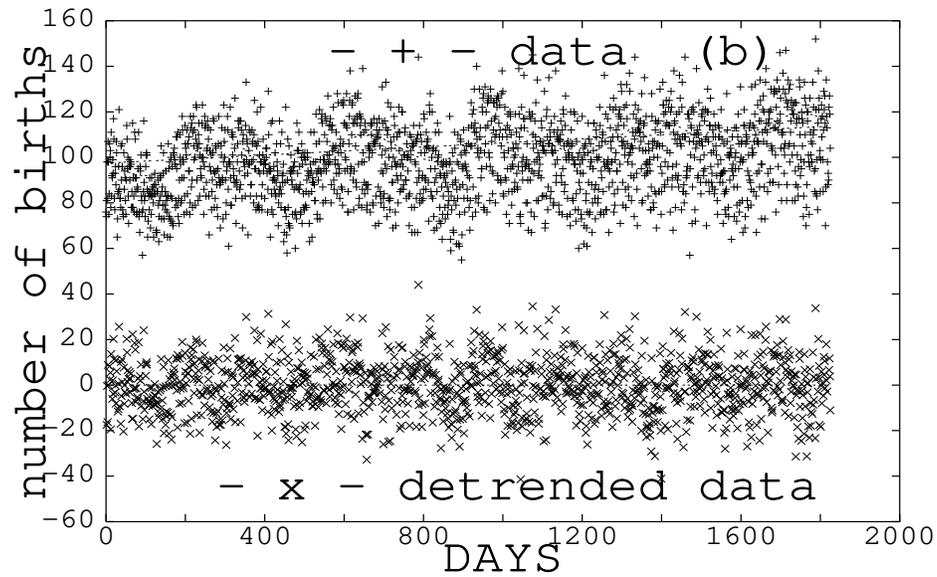, height=12cm,width=8cm,angle=-90}
\caption{Number of births (``+" symbol) to (a) married  and (b) unmarried  teenagers from January 1, 1994 through December 31, 1998. The ``x" symbol indicate the data detrended of the non-working days and of the linear and of the seasonal trends.  }
\end{figure}

\newpage
\begin{figure}[h]
\epsfig{file=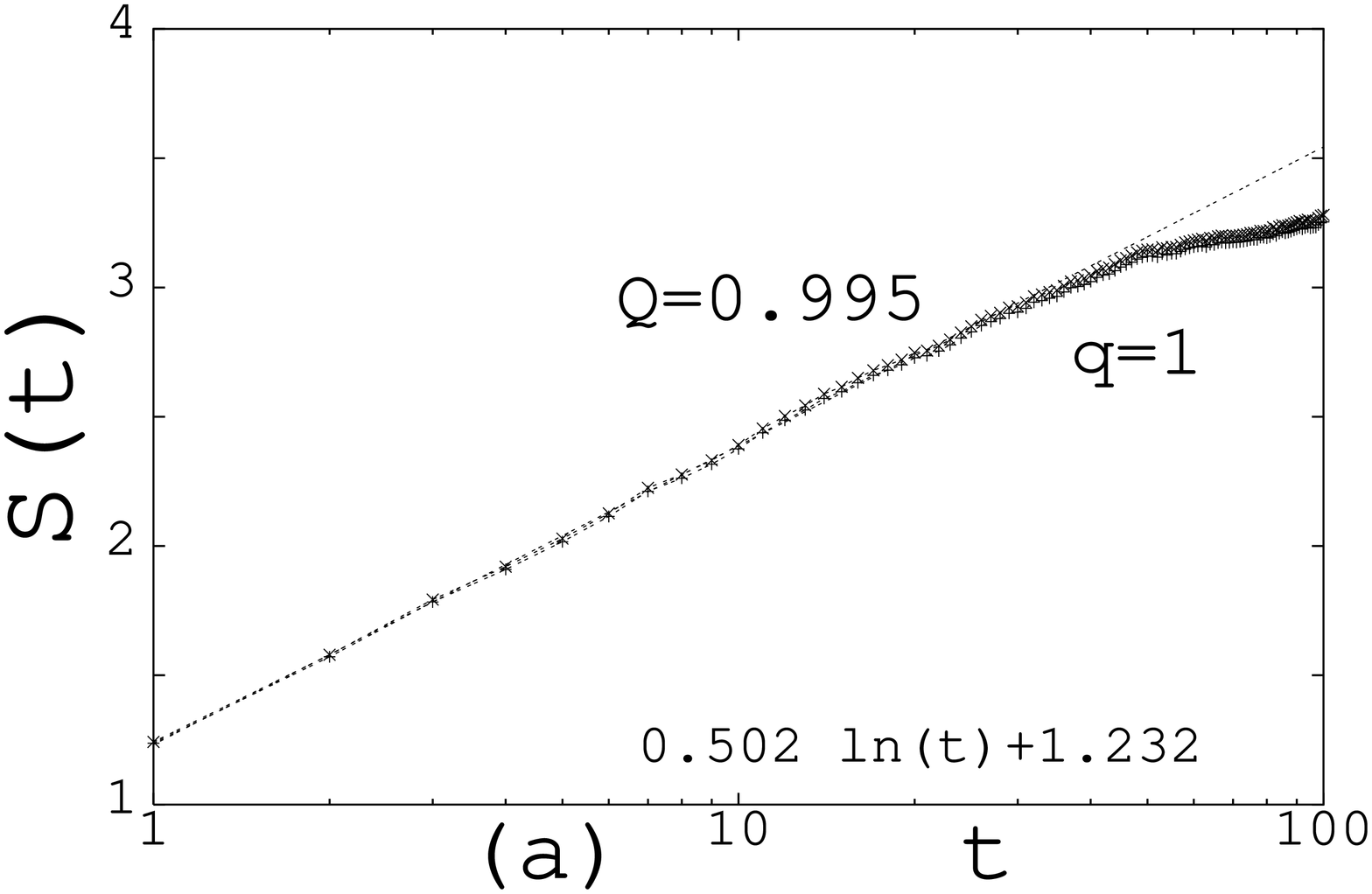, height=7cm,width=5cm,angle=-90}
\epsfig{file=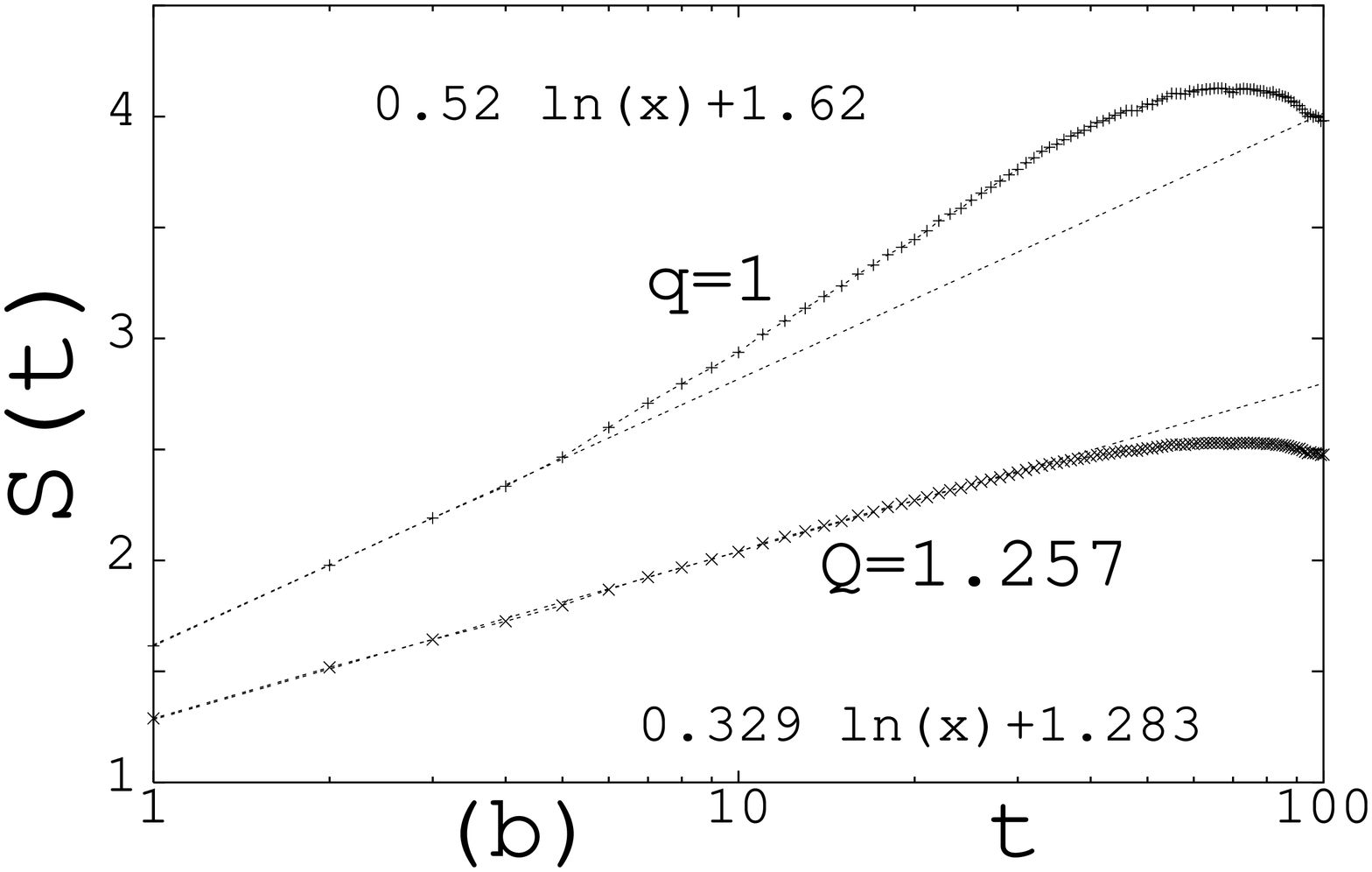, height=7cm,width=5cm,angle=-90}
\caption{Non-extensive DEA for (a) married  and (b) unmarried  teenagers. The symbol ``+" indicate the curves for $q=1$. The married teens Q=0.995 and the unmarried teens Q=1.257 curves are indicated by the symbol ``x". The fitting straight lines are shown as well.  }
\end{figure}

\begin{figure}[h]
\epsfig{file=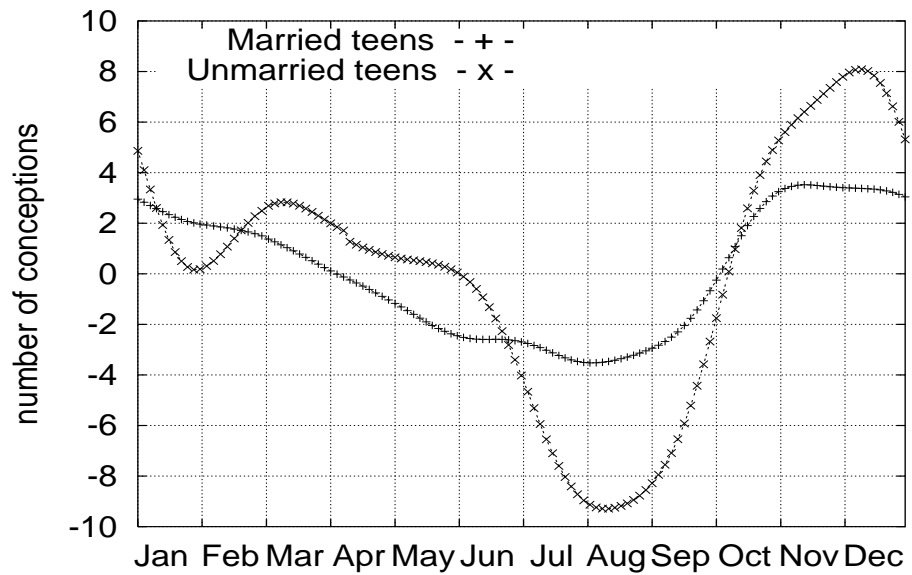, height=13cm,width=8cm,angle=-90}
\caption{Approximated daily number of  conceptions which result in birth relative to the annual mean  value in married (symbol ``+") and unmarried (symbol ``x") teenagers. The standard errors are 2  births against 2 weeks. }
\end{figure}

\end{document}